\documentclass [11pt,a4paper,notitlepage]{article}

\usepackage[english]{babel}
\usepackage[cp1250]{inputenc}
\usepackage[T1]{fontenc}
\usepackage{graphicx,epsfig}
\usepackage{amsmath}
\usepackage{amssymb}

\newcommand{\TeV}{\,\mathrm{TeV}}
\newcommand{\GeV}{\,\mathrm{GeV}}

\begin{document}

\bibliographystyle{plane}

\begin{flushright}
IFT-10/2010\\
SACLAY-T10/094
\end{flushright}

\vskip 8pt

\begin{center}
{\bf \LARGE{Higgs as a pseudo-Goldstone boson, the mu problem}}
\vskip .3cm
{\bf \LARGE{and gauge-mediated supersymmetry breaking}}
\end{center}

\vskip 24pt

\begin{center}
{\bf Anna Kami\'{n}ska$^{\, a}$ and St\'ephane Lavignac$^{\, b}$}
\end{center}


\begin{center}

\centerline{$^{a}$\textit{Institute of Theoretical Physics, Faculty of Physics,}}
\centerline{\textit{University of Warsaw, Ho{\.z}a 69, Warsaw, Poland}}
\centerline{$^{b}$\textit{Institut de Physique Th\'{e}orique, CEA-Saclay,}}
\centerline{\textit{F-91191 Gif-sur-Yvette Cedex, France\,\footnote{Laboratoire de la Direction
des Sciences de la Mati\`ere du Commissariat \`a l'Energie Atomique
et Unit\'e de Recherche Associ\'ee au CNRS (URA 2306).}}}

\end{center}

\vskip 13pt

\begin{abstract}
We study the interplay between the spontaneous
breaking of a global symmetry of the Higgs sector and gauge-mediated
supersymmetry breaking, in the framework of a supersymmetric model
with global $SU(3)$ symmetry.
In addition to solving the supersymmetric flavour problem and
alleviating the little hierarchy problem,
this scenario automatically
triggers the breaking of the global symmetry
and provides an elegant solution to the
$\mu$/$B\mu$ problem of gauge mediation.
We study in detail the processes of global symmetry and electroweak
symmetry breaking, including the contributions of the top/stop
and gauge-Higgs sectors to the one-loop effective potential
of the pseudo-Goldstone Higgs boson.
While the joint effect of supersymmetry and of the global symmetry
allows in principle the electroweak symmetry to be broken
with little fine-tuning, the simplest version of the model fails to bring
the Higgs mass above the LEP bound due to a suppressed
tree-level quartic coupling. To cure this problem, we consider
the possibility of additional $SU(3)$-breaking contributions to 
the Higgs potential, which results in a moderate fine-tuning.
The model predicts a rather low messenger scale, a small $\tan\beta$
value, a light Higgs boson with Standard Model-like properties,
and heavy higgsinos.
\end{abstract}


\section{Introduction}

Among the proposed extensions of the Standard Model,
supersymmetry is one of the most attractive from a theoretical
point of view, in particular because it automatically solves the
hierarchy problem. However, the lack of experimental evidence
put strong constraints on supersymmetric models such as the Minimal
Supersymmetric Standard Model (MSSM).
The fact that no superpartner has been discovered so far implies
a (at least in part) heavy supersymmetric spectrum, which exacerbates
the ``little hierarchy'' problem associated with the LEP bound on the
Higgs mass and increases the level of fine-tuning in the Higgs potential.
Furthermore, the absence of any significant deviation from the
Standard Model predictions in flavour physics places strong restrictions
on the generational structure of squark and slepton masses.
Gauge mediation~\cite{GMSB} offers a natural solution to this problem:
supersymmetry breaking is communicated to the observable sector
by gauge interactions, and is therefore automatically flavour blind.
On the other hand, gauge mediation suffers from the so-called
$\mu$/$B \mu$ problem~\cite{Dvali96},
i.e. the fact that the $\mu$ and $B\mu$ parameters of the MSSM
are typically generated at the same loop order, leading to
$B \mu \gg |\mu|^2$, which is inconsistent with natural electroweak
symmetry breaking.

Both problems --~the little hierarchy problem and the $\mu$/$B\mu$
problem of gauge mediation~-- may actually have a common solution
in terms of pseudo-Goldstone bosons. In Ref.~\cite{Dvali96},
a mechanism involving additional singlet superfields was proposed
to generate the $\mu$ and $B\mu$
parameters at the one- and two-loop levels, respectively, leading to
the order-of-magnitude relation $B\mu \sim |\mu|^2$.
It was pointed out that this relation has a pseudo-Goldstone interpretation:
in some limit where the Higgs superpotential becomes invariant under
a global $SU(3)$ symmetry, one combination of the two Higgs doublets
remains massless after spontaneous breaking of this symmetry
due to the relation $B\mu = |\mu|^2$
(in which the soft Higgs mass parameters have been omitted).
Regarding the little hierarchy problem, it is well known that it can
be alleviated if the Higgs boson arises as the pseudo-Goldstone
boson of some spontaneously broken approximate global symmetry,
a scenario known as little Higgs~\cite{little_Higgs}.
It was shown in Refs.~\cite{Chankowski04,Berezhiani05,Roy05,Csaki05}
that the combination of supersymmetry and of a global symmetry leads
to an improved protection
of the Higgs potential.
Namely, the logarithmic dependence of the MSSM Higgs mass
parameter on the cut-off scale is replaced by a dependence
on the scale of spontaneous global symmetry breaking, thus reducing
the need for a fine-tuning. Explicit realizations of this idea of double
protection~\cite{Berezhiani05} of the Higgs potential by supersymmetry
and by a global symmetry have shown that a fine-tuning smaller than
10\% can be achieved with squark masses around $1\TeV$.

In this paper, we revisit these issues by combining gauge-mediated
supersymmetry breaking with a spontaneously broken global
symmetry in the Higgs sector.
We consider a simple model with a global $SU(3)$ symmetry,
which is essentially a gauge-mediated version of the model of
Ref.~\cite{Berezhiani05},
and study it in detail.
We pay particular attention to the spontaneous breaking of the global
symmetry, and check whether the $SU(3)$-breaking minimum is the
global minimum of the scalar potential.
A key role in the process of global symmetry breaking is played by
the tadpole of an $SU(3)$ singlet field, which is generated by loops
of messenger fields together with the singlet soft terms.
This tadpole also triggers the generation of the $\mu$ and $B\mu$
parameters; however they do not contribute to the potential
of the lightest Higgs doublet due to its (pseudo-)\newline Goldstone boson
nature. As a result, the higgsinos and the non-Goldstone Higgs
scalars can be made heavy with masses of order $|\mu| \gg M_Z$
without introducing a strong fine-tuning in the Higgs potential.
We then compute the one-loop effective potential of the
pseudo-Goldstone Higgs doublet and 
study electroweak symmetry breaking. While the corrections
to the pseudo-Goldstone
mass parameter are efficiently controlled by
the joint effect of supersymmetry and of the global symmetry,
allowing in principle the electroweak symmetry to be broken
with little fine-tuning, the specific model studied in this paper fails to bring
the Higgs mass above the LEP bound due to a suppressed
tree-level quartic coupling. To cure this problem, we consider
the possibility of additional $SU(3)$-breaking contributions to 
the Higgs potential, and estimate the resulting fine-tuning.

The paper is organized as follows. In Section~\ref{sec:model}, we
describe the model and discuss the generation of the soft
supersymmetry breaking terms via gauge mediation.
In Section~\ref{sec:SU3_breaking}, we study the spontaneous
symmetry breaking of the global $SU(3)$ symmetry and find a
region of the parameter space in which the desired vacuum is indeed
the global minimum of the scalar potential. We then show how
the $\mu$/$B\mu$ problem is solved by the global symmetry.
Section~\ref{sec:EWSB}
deals with electroweak symmetry breaking, the Higgs mass and
fine-tuning. Finally, we give our conclusions in
Section~\ref{sec:conclusions}.

\section{The model}
\label{sec:model}

The model we study in this letter is based on a supersymmetric
version~\cite{Berezhiani05} of the ``simplest little Higgs model'' of
Ref.~\cite{Schmaltz04}. We describe its general structure below,
before discussing the generation of the soft supersymmetry breaking
terms via gauge mediation.

In order to realize the Higgs as pseudo-Goldstone boson idea, a global
$SU(3)$ symmetry spontaneously broken at the scale $f \sim 1\TeV$ is
imposed on the Higgs sector. The MSSM Higgs doublets $H_d$ and
$H_u$ are extended to global $SU(3)$ (anti-)triplets:
\begin{equation}
{\cal H}_{d}\, = \left(
\begin{array}{c}%
H_{d} \\
S_{d}
\end{array}
\right)\, \in\, {\bf 3}\ ,\ \ \ \ \ \ 
{\cal H}^T_{u}\, = \left(
\begin{array}{c}%
i \sigma^2 H_{u} \\
S_{u}
\end{array}
\right)\, \in\, {\bf \bar 3}\ ,
\label{HS}
\end{equation}
where $S_u$ and $S_d$ are electroweak singlets. Similarly, all matter
fields are extended to $SU(3)$ multiplets.
The global symmetry is spontaneously broken by the VEVs
of $S_u$ and $S_d$. The associated Goldstone boson, which is
identified with the Standard Model Higgs boson $H$, is a linear
combination of the $SU(2)_L$ doublets $H_d$ and $i \sigma^2 H^*_u$.
By construction, the tree-level Higgs potential does not contain
a mass term for $H$. However, the global $SU(3)$ symmetry of the
Higgs sector is not a symmetry of the full Lagrangian: it is violated explicitly 
by the Yukawa and gauge interactions of the MSSM. These induce a
one-loop potential for $H$ which, due to the combined effect of
supersymmetry and of the approximate global $SU(3)$ symmetry,
has no logarithmic dependence on the ultraviolet
cut-off~\cite{Berezhiani05}. 
Thanks to this softening of the radiative corrections,
the LEP Higgs mass bound can be satisfied with less fine-tuning
than in the MSSM.

In order for this double protection mechanism to be operative,
the gauge symmetry must be compatible with the global symmetry
in the ultraviolet. To this end,
the electroweak gauge symmetry
$SU(2)_L \times U(1)_Y$ is extended to $SU(3)_W \times U(1)_X$,
where $Y = X - T^8 / \sqrt{3}\,$.
The breaking of the extended gauge group is achieved at some higher
energy scale $F \gg f$ by two additional Higgs (anti-)triplets $\Phi_D$
and $\Phi_U$.
In this way, the masses of the heavy gauge bosons are
unrelated to the global symmetry breaking scale $f$, and experimental
limits on them do not constrain it. It is then possible to choose $f$
around the TeV scale, so as to minimize the fine-tuning in the
Higgs potential, without running into conflict with precision
electroweak data~\cite{Berezhiani05}.

The details of the model are presented below.

\subsection{The $SU(3)$-symmetric Higgs sector}

The Higgs sector has a global $SU(3)_{1}\times{}SU(3)_{2}$ symmetry
whose diagonal subgroup is the $SU(3)_{W}$ gauge symmetry. It contains
the following Higgs multiplets:
\begin{itemize}
\item $\Phi_{D}$ and $\Phi_{U}$, transforming as $\bf 3$ and
$\bf \bar{3}$ of $SU(3)_{1}$,
\item ${\cal H}_{d}$ and ${\cal H}_{u}$, transforming as $\bf 3$ and
$\bf \bar{3}$ of $SU(3)_{2}$,
\item two $SU(3)_{1}\times{}SU(3)_{2}$ singlets $N$ and $N'$.
\end{itemize}
Under $SU(3)_W \times U(1)_X$, $\Phi_{D}$ and ${\cal H}_{d}$
($\Phi_{U}$  and ${\cal H}_{u}$) have quantum numbers ${\bf 3}_{-1/3}$
(${\bf \bar 3}_{+1/3}$), while $N$ and $N'$ are singlets. The MSSM
Higgs doublets $H_d$ and $H_u$ are embedded in ${\cal H}_{d}$
and ${\cal H}_{u}$ as indicated in Eq.~(\ref{HS}). The
$SU(3)_{1}\times{}SU(3)_{2}$ symmetric Higgs superpotential is
chosen to be:
\begin{equation}
W_{\rm Higgs}\, =\, \lambda'N'\left(\Phi_{U}\Phi_{D}-\frac{F^2}{2}\right)
  +\lambda{}N{\cal H}_{u}{\cal H}_{d}+\frac{\kappa}{3}N^{3}\ ,
\label{eq:W_Higgs}
\end{equation}
where the last two terms are reminiscent of the NMSSM~\cite{NMSSM}.
The role of the $N{\cal H}_{u}{\cal H}_{d}$ coupling is to induce
the breaking of the global $SU(3)_2$ symmetry once the singlet
field $N$ acquires a VEV.
This coupling is also responsible, as in
the NMSSM, for the generation of the $\mu$ and $B\mu$ parameter
through the VEVs of the scalar and F-term components of $N$.
The last term is crucial to avoid a runaway of the tree-level scalar
potential in the $N$ direction.

The superpotential~(\ref{eq:W_Higgs}) leads to the spontaneous
breaking of the global $SU(3)_1$ symmetry, together with the gauge
symmetry breaking $SU(3)_W \times U(1)_X \to SU(2)_L \times U(1)_Y$:
\begin{equation}
\left\langle{}\Phi_{D}\right\rangle=\left(
\begin{array}{clrr}%
0 \\
0 \\
F_{D}
\end{array}
\right),\ \ \ \ \ \ 
\left\langle{}\Phi_{U}\right\rangle=\left(
\begin{array}{clrr}%
0 & 0 & F_{U}
\end{array}
\right) ,
\end{equation}
with $F_U = F_D = F/ \sqrt{2}$ in the supersymmetric limit
(soft terms will shift the VEVs of $\Phi_D$ and $\Phi_U$ by an amount
${\cal O} (m^2_{\rm soft} / F)$). 
The spontaneous breaking of the global $SU(3)_2$ symmetry
is triggered by a tadpole term in the singlet field $N$, whose origin
will be discussed in Section~\ref{subsec:soft}:
\begin{equation}
\left\langle{}{\cal H}_{d}\right\rangle=\left(
\begin{array}{clrr}%
0 \\
0 \\
f\cos\beta
\end{array}
\right),\ \ \ \ \ \ 
\left\langle{}{\cal H}_{u}\right\rangle=\left(
\begin{array}{clrr}%
0 & 0 & f\sin\beta
\end{array}
\right) ,
\end{equation}
where we have defined\footnote{This notation, which is reminiscent
of the one used in the MSSM for the ratio of the two Higgs doublet
VEVs, is motivated by the fact that the
pseudo-Goldstone boson is given by the same linear combination
$H \simeq \cos \beta\, H_d + \sin \beta\, (i \sigma^2 H^*_u)$
as the lightest MSSM Higgs boson in the decoupling
regime~\cite{Haber95}.} $\tan \beta \equiv \langle S_u \rangle / \langle S_d \rangle$.

One can take advantage of the hierarchy $F \gg f$ to integrate out at
the scale $F$ the heavy components of the chiral superfields $\Phi_D$,
$\Phi_U$ and $N'$, as well as the heavy gauge supermultiplets living
in the coset $SU(3)_{W}\times U(1)_{X} / SU(2)_{L}\times U(1)_{Y}$.
The resulting effective field theory is then used to study the breaking
of the global $SU(3)_2$ symmetry and of the electroweak symmetry.

\subsection{The top quark sector}

Like the Higgs fields, the matter fields of the MSSM must be extended
to $SU(3)$ multiplets. Since we are mostly interested in electroweak
symmetry breaking, we only need to consider the top/stop sector,
which gives the dominant contribution to the one-loop effective
potential\footnote{As we are going to see, $\tan \beta$ turns out to be
small in this model, so that the bottom/sbottom contribution to the
one-loop effective potential can be neglected.}. For definiteness,
we make the same choice as Ref.~\cite{Berezhiani05} for the
representations of the top quark superfields and for their couplings to the
Higgs superfields (see e.g. Refs.~\cite{Bellazzini0906,Bellazzini0910}
for alternative choices):
\begin{equation}
W_{\rm top}\, =\, y_{1}\Phi_{U}\Psi_{Q}T^{c}+y_{2}{\cal H}_{u}\Psi_{Q}t^{c} ,
\label{eq:W_top}
\end{equation}
where $\Psi_{Q} = \left(Q^T\! ,\, T\right)^{T} = \left((t, b), T\right)^{T}$
is an $SU(3)_W$ triplet, while $t^{c}$ and $T^{c}$ are $SU(3)_W$
singlets (obviously, a second singlet is necessary to render both
the top quark and its $SU(3)_W$ partner $T$ massive). Below the scale $F$, the first coupling in Eq.~(\ref{eq:W_top}) is replaced
by the effective mass term $y_{1} FTT^{c}$.
The simultaneous presence of the two terms violates the global $SU(3)_2$ symmetry; hence all $SU(3)_2$-breaking
effects from the top/stop sector will be proportional to $y_1 y_2$.

\subsection{Soft supersymmetry breaking terms}
\label{subsec:soft}

So far the model described above is identical to the one of
Ref.~\cite{Berezhiani05}. The difference lies in the soft supersymmetry
breaking terms, which in our model are calculable in terms of a few
parameters. Namely,
we assume that supersymmetry is broken in a secluded sector and
communicated to the observable sector via gauge interactions. As is
customary, we parameterize supersymmetry breaking by a gauge-singlet
spurion superfield $X$ and couple it to a vector-like pair of chiral
messenger superfields ($\Phi$, $\bar \Phi$), which we choose to be
in the representation $\left({\bf \bar 3}, {\bf 1}, \frac{1}{3}\right)
\oplus \left({\bf 1}, {\bf 3}, -\frac{1}{3}\right)$
of $SU(3)_{C}\times{}SU(3)_{W}\times{}U(1)_{X}$ and its conjugate.
In order to generate soft terms for the singlet superfield $N$, we also
introduce a coupling $N \bar \Phi \Phi$:
\begin{equation}
W_{\rm mess}\, =\, X\bar{\Phi}\Phi + \xi N\bar{\Phi}\Phi\, , \qquad
\langle X \rangle\, =\, M+\theta^{2}F_{X}\, .
\label{eq:W_mess}
\end{equation}
The soft supersymmetry breaking terms for the gauginos and
gauge non-singlet scalars are generated by the standard
messenger loops~\cite{GMSB}, and are schematically given by
(the explicit formulae can be found in Appendix~\ref{app:soft_terms}):
\begin{equation}
m_{\rm gaugino}\, \sim\, \frac{\alpha}{4\pi}\, \Lambda\, , \quad
m^{2}_{\rm scalar}\, \sim\, \left(\frac{\alpha}{4\pi}\right)^{2}\Lambda^{2}\, ,
\qquad \Lambda\equiv\frac{F_{X}}{M}\, .
\end{equation}
These expressions are valid at the messenger scale $M$, with
$\alpha = g^2(M)/4\pi$, where $g(\mu)$ is the relevant running gauge
coupling. The $A$-terms associated with the Yukawa couplings of the
top sector, $A_{y_1}$ and $A_{y_2}$, vanish at the messenger scale
and are generated at lower scales by renormalization group running. Due to the direct coupling between $N$ and the messenger fields,
soft terms for the gauge-singlet superfield $N$ are also generated
(by contrast, the soft terms for the singlet $N'$ vanish). Using the wave-function renormalization technique of Ref.~\cite{Giudice97},
we find:
\begin{equation}
A_{\lambda}\, =\, \frac{A_{\kappa}}{3}\, =\, - \frac{6\, \xi^2}{16\pi^2}\, \Lambda\, ,
\label{eq:A_lambda}
\end{equation}
\begin{equation}
m_{N}^{2}\, =\, \frac{1}{(16\pi^2)^2} \left( 48\, \xi^{4} -24\, \kappa^{2}\xi^{2}
-16\, g_{C}^{2}\xi^{2} -16\, g_{W}^{2}\xi^{2}-\frac{8}{3}\, g_{X}^{2}\xi^{2} \right)
\Lambda^{2}\, ,
\label{eq:m2_N}
\end{equation}
where $g_C$, $g_W$ and $g_X$ are the $SU(3)_C$, $SU(3)_W$
and $U(1)_X$ gauge couplings, respectively.
Note that $m^2_N < 0$ as soon as $\xi \lesssim g_W$. A negative
contribution to the soft masses of the Higgs triplets ${\cal H}_d$ and
${\cal H}_u$ is also induced by the $\xi$ coupling, on top of the standard
(positive) gauge mediation contribution:
\begin{equation}
m_{u}^{2}\, =\, m_{d}^2\, =\, \frac{1}{(16 \pi^2)^2} \left( \frac{8}{3}\, g_{W}^{4}
+\frac{8}{27}\, g_{X}^{4} -6\, \lambda^{2}\xi^{2} \right) \Lambda^{2}\, .
\label{eq:m2_u}
\end{equation}
Last but not least, the presence of a direct coupling between the singlet
$N$ and the messenger superfields also induces a tadpole in the
scalar potential~\cite{Ellwanger95,Dvali96,Ellwanger08}:
\begin{equation}
V_{\rm tad}\, =\, m^{3}N+ \mbox{h.c.}\, , \ \ \ \ \ \
m^{3}\, =\, \frac{6\, \xi}{16\pi^{2}}\, \Lambda^{2} M\, ,
\label{eq:tadpole}
\end{equation}
which plays an essential role in the breaking of the global $SU(3)_2$
symmetry, as well as an effective tadpole term in the
superpotential~\cite{Ellwanger95,Ellwanger08}:
\begin{equation}
W_{\rm tad}\, =\, M_{N}^{2}N\, ,\ \ \ \ \ \ \
M_{N}^{2}\, \sim\, \frac{6\, \xi}{16\pi^{2}}\, F^{\ast}_{X}\, .
\label{eq:W_tadpole}
\end{equation}

Let us note in passing that we could have avoided the generation of
a tadpole for $N$ by introducing a second pair of messenger fields
with the following superpotential couplings~\cite{Giudice97,Delgado08}:
\begin{equation}
W_{\rm mess}\, =\, X\bar{\Phi}_1\Phi_1 + X\bar{\Phi}_2\Phi_2
+ \xi N\bar{\Phi}_1\Phi_2\, .
\end{equation}
Since $X$ and $N$ couple to different combinations
$\Phi_i \bar \Phi_j$, no tadpole arises at one loop and  the breaking
of the global $SU(3)_2$ symmetry is triggered by the singlet soft terms
(with $m^2_N < 0$ for $\xi \lesssim g_W$).
In this letter, we choose to stick to the minimal case involving a single pair
of messenger fields, with the superpotential~(\ref{eq:W_mess}).

\section{Spontaneous breaking of the global $SU(3)_2$ symmetry}
\label{sec:SU3_breaking}

In order to proceed with the analysis of the global $SU(3)_{2}$ symmetry
breaking, it is convenient to integrate out the fields that acquire a mass
of order $F$ when the gauge symmetry $SU(3)_W \times U(1)_X$ breaks
down to $SU(2)_L \times U(1)_Y$. One is then left with the following 
tree-level potential for the fields $N$, ${\cal H}_{d}$
and ${\cal H}_{u}$, valid for energy scales $E \ll F$:
\begin{equation}
V_{\rm Higgs}\, =\, V_{F}+V_{D}+V_{\rm soft}+\delta{}V_{\rm soft}
+V_{\rm tad}\, ,
\label{eq:V_Higgs}
\end{equation}
\begin{equation}
V_{F}\, =\, \left|\lambda{\cal H}_{u}{\cal H}_{d}+\kappa{}N^{2}+M^2_N\right|^{2}
+\left|\lambda\right|^{2}\left|{}N\right|^{2}\left(\left|{\cal H}_{u}\right|^{2}
+\left|{\cal H}_{d}\right|^{2}\right) ,
\label{eq:V_F}
\end{equation}
\begin{equation}
V_{D}\, =\ \frac{g^2}{8}\, \sum_{i}\left(H_{u}^{\dag} \sigma^{i}H_{u}
+H_{d}^{\dag}\sigma^{i}H_{d}\right)^{2}
+\frac{g'^2}{8}\left(\left|{}H_{u}\right|^{2}-\left|{}H_{d}\right|^{2}\right)^{2} ,
\label{eq:V_D}
\end{equation}
\begin{equation}
V_{\rm soft}\, =\, {}m_{N}^{2}\left|N\right|^{2}
+m_{u}^{2}\left|{\cal H}_{u}\right|^{2}+m_{d}^{2}\left|{\cal H}_{d}\right|^{2}
+\left(\lambda{}A_{\lambda}N{\cal H}_{u}{\cal H}_{d}
+\frac{\kappa}{3}A_{\kappa}N^{3}+\mbox{h.c.}\right) ,
\label{eq:V_soft}
\end{equation}
\begin{equation}
\delta{}V_{\rm soft}\, =\, \left(m^{2}_{D}-m_{U}^{2}\right)
\left[\frac{9-21 t_{W}^{2}+4t^{4}_W}{36}
\left(H_{u}^{\dag}H_{u}-H_{d}^{\dag}H_{d}\right)
+\frac{1}{2}\left(S^{\ast}_{u}S_{u}-S_{d}^{\ast}S_{d}\right)\right] ,
\label{eq:deltaV_soft}
\end{equation}
\begin{equation}
V_{\rm tad}\, =\, m^{3}N+ \mbox{h.c.}\, ,
\label{Vtad}
\end{equation}
where $g$ and $g'$ are the $SU(2)_L$ and $U(1)_Y$ gauge
couplings\footnote{The matching conditions between the
$SU(3)_W \times U(1)_X$ and $SU(2)_L \times U(1)_Y$ gauge
couplings at the scale $F$ read $g = g_W$ and
$g' = g_W g_X / \sqrt{g^2_W + g^2_X/3}$.}, $t_W \equiv g'/g$ and
the expressions for the soft terms $m^2_N$, $m^2_u$, $m^2_d$,
$A_\lambda$, $A_\kappa$ and tadpole parameters $m^3$, $M^2_N$
have been given before. $V_D$ contains the $SU(2)_L \times U(1)_Y$
$D$-terms and breaks the global $SU(3)_2$ symmetry.
The term $\delta{}V_{\rm soft}$, which is a residue of integrating out
the heavy gauge supermultiplets~\cite{nondec_Dterms}, also breaks
$SU(3)_2$; it is proportional to $m^2_D -m^2_U$, the difference
between the soft masses squared of the $SU(3)_1$ Higgs triplets
$\Phi_D$ and $\Phi_U$~\cite{Berezhiani05,Roy05}. This term is
potentially dangerous because it gives a tree-level mass to
the Higgs boson, thus spoiling its pseudo-Goldstone nature.
However, since $\Phi_D$ and $\Phi_U$
are in conjugate representations, they have equal soft masses
at the messenger scale and the splitting
$m^2_D -m^2_U$ is generated by the running between $M$ and $F$; hence it
is expected to be small. Indeed, numerical calculations show that
the effect of $\delta V_{\rm soft}$ on the dynamics of $SU(3)_2$ breaking
and on the value of the Higgs mass is negligible.

The tadpole~(\ref{Vtad}) triggers a VEV $v_N \equiv \langle N \rangle
\sim m \sim \left( 6 \xi M m^2_{\rm soft} / \alpha^2 \right)^{1/3}$,
together with $f \sim m$. On the other hand, $f \lesssim 1 \TeV$ is
needed in order not to spoil the pseudo-Goldstone nature of the Higgs
boson. This points towards a rather small value of $\xi$, of the order
$\xi \lesssim 10^{-3} \alpha^2 (1 \TeV / m_{\rm soft})^2 (100 \TeV / M)$.
In practice this means that the soft terms $A_{\lambda}$, $A_{\kappa}$
and $m^2_N$, as well as the negative contribution to $m^2_u$
and $m^2_d$, are strongly suppressed and can be neglected in the
minimization of the scalar potential. As for the superpotential tadpole
term, which is of order $M^2_N \sim (6\, \xi/4\pi) M m_{\rm soft} / \alpha$,
its only effect is to shift $v_N$ by a relative amount $M^2_N/m^2 \sim
(\alpha / 4\pi)\, m/m_{\rm soft} \lesssim  (\alpha / 4\pi) (1 \TeV / m_{\rm soft})$,
and we will omit it in the following analytical considerations.
Nevertheless all these parameters are included in our numerical
computations.

While the VEV of $N$ is stabilized by the superpotential term
$\kappa N^{3}/3$,
there is no term in $V$ to stabilize a VEV
of $S_u$ ($S_d$) triggered by a negative $m^2_u$ ($m^2_d$). Since
$m^2_u$ is driven negative by the renormalization group running 
between the messenger scale $M \gtrsim 100 \GeV$ and the gauge
symmetry breaking scale $F$,
this leads to a runaway in the direction
$\langle N \rangle = \langle S_d \rangle = 0$,
$|\langle S_u \rangle| \to \infty$, which we discuss in the next subsection.

\subsection{Global $SU(3)_{2}$ symmetry breaking: analytical discussion}

Let us first ignore the runaway direction and minimize the scalar
potential for $v_{N} \equiv \left\langle N\right\rangle \neq 0$.
We have argued in the previous subsection that the soft
terms associated with $N$ as well as the superpotential tadpole
parameter $M^2_N$ can be neglected to a good approximation. Furthermore, for reasons that will become clear later, a tadpole parameter
$m$ somewhat larger than the scale of MSSM soft terms is needed in order for the proper symmetry
breaking vacuum to be the global minimum of the scalar potential.
This leads to the prediction of a small $\tan \beta$, since the
minimization conditions give:
\begin{equation}
\tan^{2}\beta\, =\, \frac{\lambda^{2}v_{N}^{2}+m_{d}^{2}}
{\lambda^{2}v_{N}^{2}+m_{u}^{2}}\ ,
\label{tan}
\end{equation}
together with $v_N \sim m$. Neglecting all soft terms (including the Higgs soft masses $m_u$
and $m_d$) in the minimization of the scalar potential, we obtain
the following approximate solution:
\begin{equation}
v_{N}^{3}\, =\, \frac{m^{3}}{2\lambda\left(2\kappa+\lambda\right)}\ , \qquad
  f\, =\, \pm\, \frac{\left[-2\left(\kappa+\lambda\right)\right]^{1/2}}
  {\lambda^{5/6}\left[2\left(2\kappa+\lambda\right)\right]^{1/3}}\ m\ , \qquad
  \tan \beta = 1\ ,
\label{app}
\end{equation}
for $\kappa+\lambda<0$.
We thus see that the spontaneous breaking of the global $SU(3)_2$
symmetry is induced by the tadpole term. To ensure that this occurs
not too far above the electroweak scale, while $m$ can be in the
multi-TeV range, some tuning between $\kappa$ and $\lambda$
is needed. If one quantifies the level of tuning by the parameter
$\epsilon > 0$, where $\kappa=-\lambda\left(1+\epsilon\right)$, then $f \simeq \mp \sqrt{\epsilon}\, m / \lambda^{2/3}$. Also,
\begin{equation}
\frac{f}{v_{N}}\, =\, \pm \sqrt{2\epsilon}\, ,
\end{equation}
implying $v_N > f$.

The problem of the runaway in the $S_{u}$ direction remains to be
discussed. As we show now, this direction is uplifted by radiative
corrections. The dominant contribution to the Coleman-Weinberg
one-loop effective potential for $S_u$,
\begin{equation}
\Delta V_{\rm 1-loop}(S_u)\, =\, \frac{1}{64\pi^{2}}\ \mbox{STr} \left[ {\cal M}^{4}(S_u)
\left( \ln \frac{{\cal M}^{2}(S_u)}{\Lambda^2} - \frac{3}{2} \right) \right] ,
\label{eq:CW}
\end{equation}
comes from the (s)top sector. The fermion mass matrix is given by:
\begin{equation}
\left(
\begin{array}{cc}%
t & T
\end{array}
\right)
\left(
\begin{array}{cc}%
0 & 0 \\
y_{2}S_{u} & y_{1}F
\end{array}
\right)
\left(
\begin{array}{cc}%
t^{c} \\
T^{c}
\end{array}
\right) ,
\end{equation}
(where we have frozen $\langle H_u \rangle = 0$), and has a single
nonzero eigenvalue corresponding to the mass of the heavy top quark,
$m_{T}=\sqrt{|y_{2}S_{u}|^{2}+|y_{1}F|^{2}}$. Neglecting the
small difference between the soft masses of the stop fields (see
Appendix~\ref{app:soft_terms}), the (s)top sector contribution to
the one-loop effective potential reads:
\begin{equation}
\Delta V_{\rm 1-loop}\, =\, \frac{3}{16\pi^{2}}\left[\left(m_{T}^{2}
  +m_{\rm stop}^{2}\right)^{2}\left(\ln\left(\frac{m_{T}^{2}
  +m_{\rm stop}^{2}}{\Lambda^{2}}\right)-\frac{3}{2}\right)
  -\, m_{T}^{4}\left(\ln\left(\frac{m_{T}^{2}}{\Lambda^{2}}\right)
  -\frac{3}{2}\right)\right] ,
\label{V1}
\end{equation}
where $m^2_{\rm stop} \equiv m^2_{\Psi_Q} = m^2_{T^c} = m^2_{t^c}$.
For large $S_u$ values, $\Delta V_{\rm 1-loop}(S_u)$ grows as
$|S_u|^2 \ln |S_u|^2$, thus curing the runaway behavior due to the
$m^2_u\, |S_u|^2$ term in the tree-level potential. However, an
unwanted minimum appears along the $S_u$ direction at the location:
\begin{equation}
S_{u,\, \rm min}^{2}\, \simeq\, \frac{F^{2}}{y_{2}^{2}}\,
\exp\left(-\frac{8\pi^{2}m_{u}^{2}}{3 y^2_2m_{\rm stop}^{2}}\right) ,
\qquad N\, \simeq\, S_d\, \simeq\, 0\, ,
\end{equation}
where we have set $\Lambda = F$ and, consistently, $m^2_u$ stands
for the running mass squared $m^2_u(F)$. The value of $S_{u,\, \rm min}$
grows exponentially as the absolute value of $m^{2}_{u}$ increases.
The value of the scalar potential at this minimum thus decreases
exponentially with $|m^2_u|$:
\begin{equation}
V_{\rm run}\left(S_{u,\, \rm min}\right)\, \simeq\,
-\, \frac{3F^{2}m_{\rm stop}^{2}}{8\pi^{2}}\,
\exp\left(-\frac{8\pi^{2}m_{u}^{2}}{3y^2_2m_{\rm stop}^{2}}\right) ,
\label{Vrun}
\end{equation}
where $V_{\rm run} (S_u) = m_{u}^{2}\, |S_{u}|^{2}
+\Delta V_{\rm 1-loop}(S_u)$.
In order to ensure that the global minimum of the scalar potential
is the one approximated by Eq.~(\ref{app}), one has to check that
$V_{\rm min} (f, v_N) < V_{\rm run} (S_{u,\, \rm min})$, where
$V_{\rm min} (f, v_N)$ is the value of the scalar potential at the
minimum~(\ref{app}):
\begin{equation}
V_{\rm min}\left(f, v_N\right)\, \simeq\, \frac{3}{2}\, m^{3}v_N\, \simeq\,
-\, \frac{3m^{4}}{2\left|2\lambda\left(2\kappa+\lambda\right)\right|^{1/3}}\ .
\end{equation}
This requirement imposes some
restrictions on the parameters of the
model, most notably on the messenger mass $M$ and on the tadpole
scale $m$ (or equivalently on $M$ and on the parameter $\xi$). In
particular, larger values of $m$ are preferred, making it necessary
to slightly tune the values of $\kappa$ and $\lambda$ in order to
maintain the $SU(3)_2$ breaking scale $f$ below $1 \TeV$ or so.
Numerical calculations show that $M \lesssim 1000 \TeV$ and
$\xi\gtrsim10^{-4}$ lead to reasonable results.

\subsection{Global $SU(3)_{2}$ symmetry breaking: numerical results}

In order to study numerically the spontaneous breaking of the global
$SU(3)_{2}$ symmetry, we first set the values of the parameters
$F$, $y_{1}$, $y_{2}$, $\kappa$, $\lambda$ and
of the various soft terms at the messenger scale and perform the appropriate
renormalization group (RG) running. The $SU(3)_W$ breaking scale
$F$ and the coupling $y_1$ cannot be chosen too large if one wants
to uphold the pseudo-Goldstone nature of the Higgs doublet, since
the $SU(3)_{2}$ violating effects in the (s)top sector are proportional
to $y_1 F$, as discussed in the next section. All the effects neglected
in the analytical discussion above (like the superpotential tadpole
term, the soft terms in the Higgs potential and the contributions from
the Higgs sector to the Coleman-Weinberg effective potential)
are taken into account numerically.
The $SU(3)_{2}$-symmetric RGEs valid between the messenger
scale $M$ and the $SU(3)_{W}$ breaking scale $F$ can be found in
Appendix~\ref{app:RGEs}, while Appendix~\ref{app:soft_terms} gives
the boundary conditions for the soft terms at the messenger scale,
calculated using the wave-function renormalization technique of
Ref.~\cite{Giudice97}.

\begin{figure}[h!]
\begin{minipage}[b]{0.47\linewidth}
\centering
\includegraphics[width=6 cm]{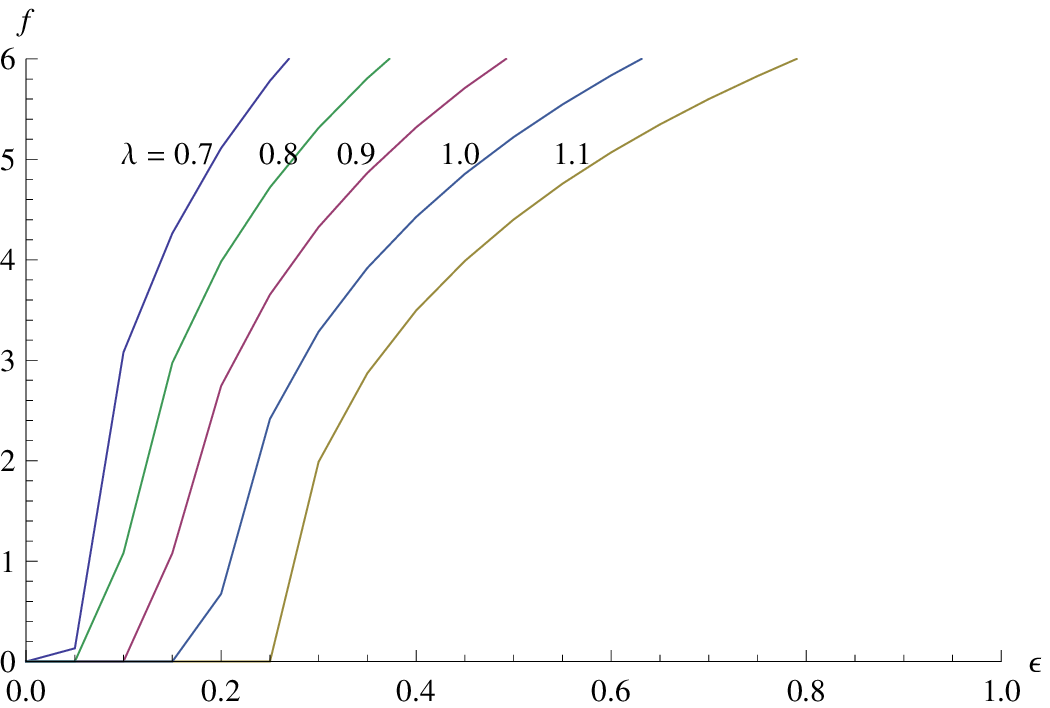}
\caption{$f \,$[TeV] as a function of $\lambda$ and $\epsilon$, \newline
for $M = 500 \TeV$, $m_{\rm soft} \sim 1 \TeV$, $y_{1}=0.1$, \newline
$F=10 \TeV$ and $\xi = 0.002$.}
\label{fig:f_xi-002}
\end{minipage}
\hspace{0.5cm}
\begin{minipage}[b]{0.47\linewidth}
\centering
\includegraphics[width=6 cm]{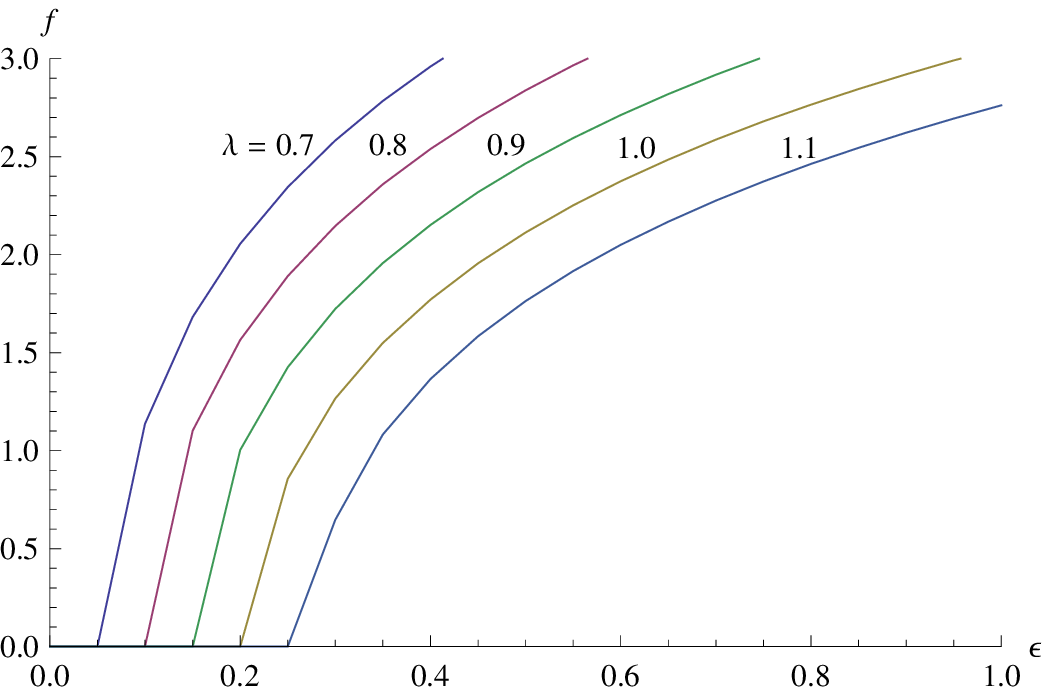}
\caption{$f\,$[TeV] as a function of $\lambda$ and $\epsilon$, \newline
for $M=500\TeV$, $m_{\rm soft} \sim 1\TeV$, $y_{1}=1$, \newline
$F=7\TeV$ and $\xi=0.00014$.}
\label{fig:f_xi-00014}
\end{minipage}
\end{figure}

The  minimization of the scalar potential at the scale $F$ leads to
the spontaneous breaking of the $SU(3)_{W} \times U(1)_X$ gauge
symmetry, together with the $SU(3)_1$ global symmetry. Then
the heavy degrees of freedom are integrated out and the breaking
of the global $SU(3)_{2}$ symmetry is studied by minimizing the
effective potential below $F$, defined as the sum of the tree-level
potential~(\ref{eq:V_Higgs}) with its parameters renormalized at
the scale $F$ and of the Coleman-Weinberg one-loop corrections 
computed with $\Lambda = F$.
The result is then confronted
with the requirement of having the proper global minimum of the
potential, with the $SU(3)_2$ breaking scale $f$ not too far above
the electroweak scale, and a correct prediction for the top quark
Yukawa coupling. The approximate proportionality between
the VEVs $v_N$, $f$ and the tadpole scale $m$, Eq.~(\ref{app}),
is confirmed by numerical calculations. A slightly more accurate set
of equations for $v_N$ and $f$ is given by:
\begin{equation}
f^{2}\, \simeq\, \frac{1}{\lambda^{2}}\left[-2\lambda\left(\kappa+\lambda\right)v_{N}^{2}-m_{u}^{2}-m_{d}^{2}\right] ,
\label{cf4}
\end{equation}
\begin{equation}
2 \lambda^2 (2\kappa + \lambda) v_{N}^{3}
+ (\kappa + \lambda) (m^2_u + m^2_d) v_N - \lambda m^3 \simeq\, 0\, ,
\label{eq:vN}
\end{equation}
and the prediction $\tan\beta \simeq 1$ holds.
Figs.~\ref{fig:f_xi-002} and~\ref{fig:f_xi-00014} illustrate the dependence of the
$SU(3)_2$ breaking scale $f$ on the parameters $\kappa$ and
$\lambda$, for a messenger mass $M = 500\TeV$, a typical soft mass
scale $m_{\rm soft} \sim 1\TeV$ and various choices for $\xi$, $y_1$
and $F$.
All coupling values in the figures are given at the messenger scale.
As shown by these plots, $f \lesssim 1\TeV$ can be achieved with
$\lambda \sim 1$ and a mild tuning between $\kappa$ and
$\lambda$, $\epsilon \sim 0.2$.
These values imply a moderate hierarchy between the $SU(3)_2$
breaking scale $f$ and the VEV $v_N$.

\subsection{The solution of the $\mu$/$B\mu$ problem}

Let us now discuss the generation of the $\mu$ and $B\mu$ terms.
After spontaneous breaking of the global $SU(3)_2$
symmetry, the doublet and singlet components of the Higgs triplets
${\cal H}_u$ and ${\cal H}_d$ no longer share the same masses
and couplings, and the quadratic part of the tree-level scalar
potential~(\ref{eq:V_Higgs}) can be rewritten as:
\begin{eqnarray}
V_{\rm quadr.} & \!\! = & \!\! |\mu|^2 \left( |H_u|^2 + |H_d|^2 \right)
+ \left( B\mu\, H_u\! \cdot\! H_d + \mbox{h.c.} \right)  \nonumber \\
&& +\, \frac{1}{2} \left( s_u\ s_d\ s_N \right) M^2_S
\left(\!\!\! \begin{array}{c} s_u \\ s_d \\ s_N  \end{array}\!\!\! \right)
+ \frac{1}{2} \left( p_u\ p_d\ p_N \right) M^2_P
\left(\!\!\! \begin{array}{c} p_u \\ p_d \\ p_N  \end{array}\!\!\! \right) ,
\label{eq:V_quadr}
\end{eqnarray}
where $H_u\! \cdot\! H_d \equiv H^T_u i \sigma^2 H_d$, 
$S_{u,d} \equiv f_{u,d} + (s_{u,d} + i p_{u,d}) / \sqrt{2}$,
$N \equiv v_N + (s_N + i p_N) / \sqrt{2}$ and all parameters
in the scalar potential are assumed to be real.
As in the NMSSM, the $\mu$ and $B\mu$ parameters are
generated by the VEVs of the scalar and F-term components
of the singlet superfield $N$ (the signs in the expressions
for $\mu$ and $B\mu$ are due to the fact that
${\cal H}_u {\cal H}_d = -H^T_u i \sigma^2 H_d + S_u S_d$):
\begin{equation}
  \mu\, =\, - \lambda v_N\, , \qquad \quad
  B\mu\, =\, \lambda F_N - \lambda A_\lambda v_N\,
  =\, - \lambda \left( \lambda f_u f_d + \kappa v^2_N + A_\lambda v_N \right) .
\end{equation}
However, there is a crucial difference with the NMSSM: here $F_N$
receives a contribution from the VEVs of $S_u$ and $S_d$, 
which transform non-trivially under the global $SU(3)_2$ symmetry.
At the minimum of the tree-level scalar potential~(\ref{eq:V_Higgs}),
these VEVs take values such that the relation
$(B\mu)^2 = (\mu^2 + m^2_u)(\mu^2 + m^2_d)$ is satisfied.
This in turns implies that the determinant of the ($H_u$, $H_d$)
mass matrix vanishes (a similar mechanism is at work in the
$SU(3)$-symmetric version of the model of Ref.~\cite{Dvali96}).
The massless combination:
\begin{equation}
 H\, =\, \sin \beta\, (i \sigma^2 H^*_u) + \cos \beta\, H_d\, ,
\label{eq:Goldstone_doublet}
\end{equation}
to be identified with the Standard Model Higgs boson,
is interpreted as a Goldstone boson of the spontaneously broken
global $SU(3)_2$ symmetry. The orthogonal combination $H'$
is heavy with a mass $m^2_{H'} = 2 \mu^2 + m^2_u + m^2_d$.
Inspection of the singlet scalar and pseudoscalar mass matrices
$M^2_S$ and $M^2_P$ show that there is another Goldstone boson,
\begin{equation}
 \eta \,= \,\sin \beta\, p_u - \cos \beta\, p_d\, .
\label{eq:Goldtsone_singlet}
\end{equation}
The other singlets
are massive with masses of order a few $\mu$, except for a lighter
one with a mass of order $\lambda f \lesssim 1 \TeV$.

If we restrict our attention to the part of the tree-level scalar potential
that depends solely on $H$, we see no dependence
on $\mu$ and $B\mu$, while the masses of the heavy states
of the Higgs sector (including the higgsinos) are of order $\mu$.
Hence, the electroweak scale is insensitive to the actual value
of the $\mu$ parameter, which is allowed to be large without
creating a strong fine-tuning in the Higgs potential\footnote{In fact,
radiative corrections induce a dependence of the one-loop effective
potential on $\mu$ and $B\mu$ (see Section~\ref{subsec:EWSB}),
but this does not represent an important source of fine-tuning.}.
This elegantly solves the $\mu$/$B\mu$ problem of gauge mediation.
The value of $\mu$ (or $v_N$) is relevant, on the other hand,
for the breaking of the global symmetry, and we have seen in the
previous subsections that a moderate hierarchy $\mu \gg f$ is needed,
with no incidence on fine-tuning by virtue of the global $SU(3)_2$
symmetry.

\section{Electroweak symmetry breaking }
\label{sec:EWSB}

We are now in a position to discuss the breaking of the electroweak
symmetry. Let us first recapitulate the identification of the light degrees
of freedom in the Higgs sector.
The spontaneous breaking of the global
$SU(3)_1 \times SU(3)_2$ symmetry leads to 10 Goldstone bosons,
5 of which disappear from the massless spectrum by virtue of the
Higgs mechanism, since the gauge symmetry $SU(3)_W \times U(1)_X$
is broken to $SU(2)_L \times U(1)_Y$ in the same process. 
The remaining 5 Goldstone bosons reside mainly in ${\cal H}_u$
and ${\cal H}_d$ in the limit $f \ll F$, and are conveniently
parameterized as\footnote{This parametrization agrees with
Eqs.~(\ref{eq:Goldstone_doublet}) and~(\ref{eq:Goldtsone_singlet})
at leading order in $1/f$.}~\cite{Berezhiani05}:
\begin{equation}
{\cal H}_{d}\, =\, f_{d}\left(
\begin{array}{c}%
\frac{H}{\left|{}H\right|}\, \sin\left(\frac{\left|{}H\right|}{f}\right) \\
e^{-\frac{i\eta}{f\sqrt{2}}}\cos\left(\frac{\left|{}H\right|}{f}\right)
\end{array}
\right),\ \ \ \ \ \ 
{\cal H}_{u}\, =\, f_{u}\left(
\begin{array}{cc}%
\frac{H^{\dag}}{\left|{}H\right|}\, \sin\left(\frac{\left|{}H\right|}{f}\right) ,  &
\, e^{\frac{i\eta}{f\sqrt{2}}}\cos\left(\frac{\left|{}H\right|}{f}\right)
\end{array}
\right) ,
\label{eq:Goldstones}
\end{equation}
where $f_{u} \equiv f\sin\beta$, $f_{d} \equiv f\cos\beta$ and
$\left|{}H\right| \equiv \sqrt{H^{\dag}H}$. All other components
of the Higgs triplets,
except for one singlet with mass of order $\lambda f$, are heavy
with masses of order a few $\mu \gg f$ and can be integrated out.
$H$ is a Standard Model-like
Higgs doublet transforming as a ${\bf 2}_{-1/2}$ of
$SU(2)_{L}\times{}U(1)_{Y}$, while $\eta$ is a singlet whose role
has been discussed in
Refs.~\cite{Berezhiani05,Bellazzini0906,Bellazzini0910}.
Being a (pseudo-)Goldstone boson of the approximate global
$SU(3)_2$ symmetry, $H$ gets its potential from $SU(3)_2$
breaking interactions. At tree level, one has:
\begin{equation}
V_{\rm tree} (H)\, =\, V_{\rm light} (H) + V_{\rm heavy} (H)\, ,
\end{equation}
where $V_{\rm light} (H)$ is the contribution of the
$SU(2)_L \times U(1)_Y$ $D$-terms:
\begin{equation}
V_{\rm light} (H)\, =\, \lambda_0 \left\{\, |H|^4
  + {\cal O} \left( \frac{|H|^6}{f^2} \right) \right\} ,
  \qquad \lambda_0\, =\, \frac{g^{2}+g'^{2}}{8}\, \cos^{2}2\beta\, ,
\label{eq:lambda_0}
\end{equation}
and $V_{\rm heavy} (H)$ is the contribution of the terms
$\delta{}V_{\rm soft}$ left over by integrating out the heavy gauge
supermultiplets at the scale $F$:
\begin{equation}
V_{\rm heavy} (H)\, =\, m^2_0
  \left\{ |H|^2 + {\cal O} \left( \frac{|H|^4}{f^2} \right)\! \right\} , \\
\quad m^2_0\, =\, \frac{9-21 t_{W}^{2}+4t^{4}_W}{36}\,
  (m^{2}_{D}-m_{U}^{2}) (- \cos 2 \beta)\, ,
\end{equation}
($m^2_0 > 0$ due to $\cos 2 \beta < 0$ and $m^2_D -m^2_U > 0$).
Since $\tan \beta \simeq 1$, both the tree-level quartic coupling
$\lambda_0$ and the mass parameter $m^2_0$ are small\footnote{As
discussed at the beginning of Section~\ref{sec:SU3_breaking},
$m^2_0$ is further suppressed by the small RG-induced difference
of heavy Higgs triplet soft masses $m^2_D -m^2_U$.
It is therefore not expected to play a significant role in the dynamics
of electroweak symmetry breaking. This is confirmed by numerical
calculations, which show that the tree-level contributions to the
Higgs potential are negligible in comparison with the one-loop corrections.}.
As we are going to see in the next subsection, one-loop corrections
induced by the large top quark Yukawa coupling generate a tachyonic
mass term in the Higgs potential and trigger electroweak symmetry
breaking. The electroweak scale $v$ is related to the VEV of the Higgs
doublet $\bar v \equiv \langle H \rangle$ by:
\begin{equation}
v\, =\, f\sin\left(\bar{v}/f\right) .
\end{equation}
%

\subsection{Electroweak symmetry breaking: analytical discussion}
\label{subsec:EWSB}

At the one-loop level, $V(H)$ receives contributions
from the Higgs couplings to the matter and gauge fields, which
explicitly break the global $SU(3)_2$ symmetry. Let us first compute
the radiative corrections induced by the top quark Yukawa coupling,
using the Coleman-Weinberg formula (Eq.~(\ref{eq:CW}) with $S_u$
replaced by $H$). The fermion mass matrix squared reads, in the
parameterization~(\ref{eq:Goldstones}):
\begin{equation}
M_{\rm top}^{\dag} M_{\rm top}\, =
\left( \begin{array}{cc}
y_{2}^{2}f_{u}^{2} & y_{1}y_{2}Ff_{u}\cos\left(\left|{}H\right|/f\right) \\
y_{1}y_{2}Ff_{u}\cos\left(\left|{}H\right|/f\right) & y_{1}^{2}F^{2}
\end{array} \right)\, ,
\label{Mt}
\end{equation}
where ${\cal L}_{\rm mass} \ni -\left(t\, \ T\right) M_{\rm top}
\left(t^{c}\, \ T^{c}\right)^T + \mbox{h.c.}\,$. The eigenstates can be
identified with the Standard Model top quark and its heavy $SU(3)$
partner, with masses:
\begin{equation}
\left( m^T_t \right)^2\, =\, \frac{1}{2}\left(y_{1}^{2}F^{2}+y_{2}^{2}f_{u}^{2}
\pm \sqrt{\left(y_{1}^{2}F^{2}+y_{2}^{2}f_{u}^{2}\right)^{2}
- 4y_{1}^{2}y_{2}^{2}F^{2}f_{u}^{2}\sin^{2}\left(\left|{}H\right|/f\right)}\, \right) .
\label{eq:m_T_t}
\end{equation}
For $|H| \ll f$, Eq.~(\ref{eq:m_T_t}) simplifies to:
\begin{equation}
m_{t}^{2}\, \simeq\, {}y_{t}^{2}\left|{}H\right|^{2}, \qquad
m_{T}^{2}\, \simeq\,{}y_{1}^{2}F^{2}+y_{2}^{2}f_{u}^{2}\ ,
\end{equation}
where
\begin{equation}
y_{t}^{2}\, =\, \frac{y_{1}^{2}y_{2}^{2}F^{2}\sin^{2}\beta}
{y_{1}^{2}F^{2}+y_{2}^{2}f_{u}^{2}}\ .
\end{equation}
Plugging these expressions into the Coleman-Weinberg formula
and neglecting the small difference between the soft masses
of the stop fields, one obtains similar expressions to the ones
of Ref.~\cite{Berezhiani05}:
\begin{equation}
\delta_{t}m_{H}^{2}\, \simeq\, -\frac{3y^2_t}{8\pi^{2}}
\left[\, m_{\rm stop}^{2}\ln\left(1+\frac{m_{T}^{2}}{m_{\rm stop}^{2}}\right)
+m_{T}^{2}\ln\left(1+\frac{m_{\rm stop}^{2}}{m_{T}^{2}}\right)\right]
\label{dm}
\end{equation}
and
\begin{equation}
\delta_{t}\lambda_H\, \simeq\, \frac{3y^4_t}{16\pi^{2}}
\left[\, \ln\left(\frac{m_{\rm stop}^{2}m_{T}^{2}}
{m_{t}^{2}\left(m_{\rm stop}^{2}+m_{T}^{2}\right)}\right)
-2\, \frac{m_{\rm stop}^{2}}{m_{T}^{2}}\,
\ln\left(1+\frac{m_{T}^{2}}{m_{\rm stop}^{2}}\right)\right.
\nonumber
\end{equation}
\begin{equation}
\hskip 3cm \left.+\, \frac{2m_{\rm stop}^{2}}{3y_{t}^{2}f^2}\,
\ln\left(\frac{m_{\rm stop}^2 + m_{T}^{2}}{m_{\rm stop}^{2}}\right)
+\frac{2m_{T}^{2}}{3y_{t}^{2}f^2}\,
\ln\left(\frac{m_{\rm stop}^{2} + m_T^2}{m_{T}^{2}}\right)\right] ,
\end{equation}
where $\delta_{t}m_{H}^{2}$ and $\delta_{t}\lambda_H$ are the
contributions of the (s)top sector to the coefficients of the quadratic
and quartic terms in the one-loop effective Higgs potential, respectively:
\begin{equation}
\Delta V_{\rm 1-loop} (H)\, =\, \delta{}m_{H}^{2}\left|{}H\right|^{2}
+ \delta\lambda_H \left|{}H\right|^{4}+ \cdots\ .
\label{ex}
\end{equation}
As required for proper electroweak symmetry breaking,
$\delta_{t}m_{H}^{2}$ is negative while $\delta_{t}\lambda_H$ is positive.
The absence of a $\ln \Lambda$-dependent piece in
$\delta_{t}m_{H}^{2}$ is a direct consequence of the double protection
of the Higgs mass by supersymmetry and by the global $SU(3)_2$
symmetry. 

Let us now consider the contributions of the gauge
interactions to $\delta m_{H}^{2}$ and $\delta \lambda_H$.
In the effective theory below the gauge symmetry breaking scale $F$,
these depend logarithmically on the cut-off scale $\Lambda$ due to
the explicit breaking of the global $SU(3)_2$ symmetry
by the $SU(2)_L \times U(1)_Y$ gauge couplings and gaugino masses.
The dominant contributions to the one-loop effective potential are
given by the approximate formula:
\begin{equation}
\Delta_{\rm gauge} V_{\rm 1-loop} (H)\, \simeq\,
-\frac{1}{64\pi^2}\ \mbox{STr} \left[ {\cal M}^{4}_{\rm gauge-Higgs} (H)\,
\ln \frac{\Lambda^{2}}{m^{2}_{\rm soft}} \right] ,
\label{eq:CW_g}
\end{equation}
where ${\cal M}^4_{\rm gauge-Higgs}$ stands for the (fourth power
of the) mass matrices of the gauge and Higgs fields, and $m_{\rm soft}$
is an average soft mass. Since the whole gauge sector is $SU(3)_2$
symmetric above the scale $F$, the logarithmic divergence is
effectively cut off at $\Lambda = F$. Let us compute~(\ref{eq:CW_g}).
Working in the approximation where the heavy
gauge and Higgs fields are integrated out at tree level, one
is left with the $SU(2)_L \times U(1)_Y$ gauge fields and with the
Higgs superfields ${\cal H}_{u}$, ${\cal H}_{d}$ and $N$. Using the
Higgs superpotential $W = \lambda{}N{\cal H}_{u}{\cal H}_{d}
+\frac{\kappa}{3}N^{3}$, the tree-level
potential~(\ref{eq:V_Higgs})--(\ref{Vtad}) and the Lagrangian terms
involving the $SU(2)_L \times U(1)_Y$ gauge fields, one derives
the mass matrices of the gauge bosons, charginos,
neutralinos, charged and neutral Higgs bosons. Neglecting
$\delta V_{\rm soft}$ as well as the soft terms that are suppressed by
the small parameter $\xi$, and assuming all parameters to be real,
one obtains:
\begin{eqnarray}
& \mbox{STr} \left[ {\cal M}^{4}_{\rm gauge-Higgs} \right]\, =\,
-\, \Big[ 4 ( 3g^2 M^2_2+g^{\prime 2} M^2_1)
+ 3 ( 3g^2+g^{\prime 2}) \mu^2 \Big]
\left( \left| H_{u}\right|^2+\left| H_{d}\right|^2\right) 
\nonumber  \\
& +\ \Big(\, \Big[ 4 ( 3g^2 M_2+g^{\prime 2} M_1 ) \mu
- ( 3g^2+g^{\prime 2} ) B\mu \Big]\, H_{u}\! \cdot\!  H_{d}\,
+\, \mbox{h.c.}\, \Big) 
\nonumber  \\
& +\ 3 ( g^2+g^{\prime 2} ) \left[ m_{u}^{2}\left| H_{u}\right|^2
+ m_{d}^{2}\left| H_{d}\right|^2\right] -\, 2g^{\prime 2}
\left[ m_{d}^{2}\left| H_{u}\right|^2+m_{u}^{2}\left| H_{d}\right|^2\right] 
\nonumber  \\
& +\, \left[ \frac{(g^2+g^{\prime 2})^2}{2} - \frac{9}{4}\, g^4
+ \frac{1}{4}\, g^{\prime 4} - \lambda^2 (g^2+g^{\prime 2}) \right]
\left( \left| H_{u}\right|^2-\left| H_{d}\right|^2\right) ^{2},
\label{eq:STrM4_gH}
\end{eqnarray}
where field-independent terms have been omitted, and $\mu= - \lambda v_{N}$, 
$B\mu=-\lambda ( \lambda f_{u} f_{d}+\kappa v_{N}^{2})$. Inserting
the parameterization~(\ref{eq:Goldstones}) into Eq.~(\ref{eq:STrM4_gH}),
one finally obtains:
\begin{eqnarray}
\delta_{g}m^{2}_H & \simeq & \left\lbrace\,
\frac{3g^2 M^2_2+g^{\prime 2} M^2_1}{16\pi^2}\,
+\, \frac{3}{64\pi^2}\, ( 3g^2+g^{\prime 2}) \mu^2\,
+\, \frac{3g^2 M_2+g^{\prime 2} M_1}{16\pi^2}\, \mu \sin 2\beta \right.
\nonumber  \\
&& -\, \frac{3g^2+g^{\prime 2}}{64\pi^{2}}\, B\mu\, \sin 2\beta\,
-\, \frac{3( g^2+g^{\prime 2})}{64\pi^{2}}\,
\Big[ m_{u}^{2}\sin^{2}\! \beta\, +\, m_{d}^{2}\cos^{2}\! \beta \Big] 
\nonumber  \\
&& + \left. \frac{g^{\prime 2}}{32\pi^{2}}\,
\Big[ m_{d}^{2}\sin^{2}\, \beta+m_{u}^{2}\cos^{2}\, \beta \Big]\,
\right\rbrace\, \ln \left( \frac{F^{2}}{m^{2}_{\rm soft}} \right)\, ,
\label{eq:deltag_m2}  \\
\delta_{g} \lambda_H & \simeq &
-\, \frac{1}{64\pi^2} \left[ \frac{(g^2+g^{\prime 2})^2}{2} - \frac{9}{4}\, g^4
+ \frac{1}{4}\, g^{\prime 4} - \lambda^2 (g^2+g^{\prime 2}) \right]
\cos^{2}2\beta\, \ln \left( \frac{F^{2}}{m^{2}_{\rm soft}}\right)
\nonumber  \\
&& -\, \frac{\delta_{g}m^{2}_H}{3f^2}\, ,
\end{eqnarray}
where we have set $\Lambda = F$. The terms enhanced
by $\mu^2$ and $\mu M_{1,2}$ dominate in $\delta_{g}m^{2}_H$, so
that $\delta_{g}m^{2}_H > 0$. Note that there is a partial cancellation
between the second and the fourth terms, due to
$B \mu \simeq -\lambda \kappa v^2_N \simeq \lambda^2 v^2_N = \mu^2$,
leaving a net contribution $( 3g^2+g^{\prime 2})\, \mu^2
\ln (F^{2} / m^{2}_{\rm soft}) / 32 \pi^2$.
Contrary to $\delta_{t} \lambda_H$, $\delta_{g} \lambda_H$ is negative,
but it is suppressed by $\cos^2 2 \beta$ (first term) and by $1/3f^2$
(second term).

The second and fourth terms in Eq.~(\ref{eq:deltag_m2}),
which due to the large value of $\mu$ (see next subsection) give
the dominant contribution from the gauge-Higgs sector to the
Coleman-Weinberg potential, have a simple renormalization group
interpretation. They arise from the different RG running,
below the $SU(3)_W$ breaking
scale $F$, of the parameters associated with the doublet and singlet
components of the Higgs triplets ${\cal H}_u$ and ${\cal H}_d$.
Indeed, below $F$, gauge interactions distinguish the doublets
$H_u$ and $H_d$ from their $SU(3)$ partners $S_u$ and $S_d$, 
and this results in different RGEs for parameters that would otherwise
be equal by virtue of the $SU(3)_2$ symmetry. One is thus led to 
``split'' the superpotential coupling $\lambda$ in the following way:
\begin{equation}
W_{\rm Higgs}\, \ni\, \lambda_{s}NS_{u}S_{d}+\lambda_{d}NH_{u}H_{d}\, ,
\label{sd}
\end{equation}
and similarly for the soft terms involving ${\cal H}_u$ or ${\cal H}_d$.
As a result, the $F$-term potential~(\ref{eq:V_F}) is modified as follows:
\begin{equation}
V_{F}\, =\, \left(\left[\lambda_{s}\cos^{2}\left(\frac{\left| H\right|}{f}\right)
+\lambda_{d}\sin^{2}\left(\frac{\left| H\right|}{f}\right)\right]\!
f^{2}\sin\beta\cos\beta+\kappa v_{N}^{2}\right)^{2}
\nonumber
\end{equation}
\begin{equation}
+\, v_{N}^{2}f^{2}\left(\lambda^{2}_{s}\cos^{2}\left(\frac{\left| H\right|}{f}\right)
+\lambda^{2}_{d}\sin^{2}\left(\frac{\left| H\right|}{f}\right)\right) ,
\label{eq:V_F_split}
\end{equation}
where we have replaced $N$ by its VEV and inserted
the parameterization~(\ref{eq:Goldstones}). Using $B\mu \simeq \mu^2$
and $\sin 2\beta \simeq 1$, this yields:
\begin{equation}
\delta_{\rm split} m^2_H\, \simeq\, \frac{\lambda_d - \lambda_s}{\lambda}\, \mu^2\, ,
\quad \delta_{\rm split} \lambda_H\, \simeq\, -\, \frac{\delta_{\rm split}m^2_H}{3 f^2}\, ,
\end{equation}
where $\lambda_d - \lambda_s / \lambda$ can be computed from
the RGEs for the ``split'' superpotential couplings given in 
Appendix~\ref{app:RGEs_split}:
\begin{equation}
\frac{\lambda_d - \lambda_s}{\lambda}\ \simeq\ \ln \frac{\lambda_d}{\lambda_s}\
\simeq\ \frac{3 g^2 + g^{\prime 2}}{32 \pi^2}\, \ln \left( \frac{F^{2}}{m^{2}_{\rm soft}}\right) ,
\end{equation}
in agreement with the second and fourth terms of Eq.~(\ref{eq:deltag_m2}).

\subsection{Electroweak symmetry breaking: numerical results}

The numerical study of electroweak symmetry breaking is done by
minimizing the Higgs potential, taking into account the contributions
mentioned in the previous subsections and relaxing the assumptions
made in the analytical discussion (in particular, the soft masses
in the stop sector are not universal but given by the formulae of
Appendix~\ref{app:soft_terms}). For simplicity, only the dominant
(s)top sector contribution and the approximate gauge
contribution~(\ref{eq:CW_g})--(\ref{eq:STrM4_gH})
have been included in the numerical
computation of the Coleman-Weiberg potential.

Let us comment on the features of the main contributions
to the Higgs potential, as revealed by the numerical calculations.
The contribution from the (s)top sector to the Coleman-Weinberg
one-loop effective potential has the desired ``mexican hat'' shape
with the minimum approximately located at $\bar{v}\approx\frac{\pi}{2}f$.
The gauge contribution is convex for small values of the Higgs VEV
with a minimum at the origin, which helps shifting the minimum of the
Higgs potential towards the correct value $\bar v \approx v = 174 \GeV$. 
However, the LEP bound on the Higgs mass
requires large corrections from the (s)top sector to the quartic
coupling $\lambda_H$ and at the same time does not
allow for large corrections from the gauge-Higgs sector (which
gives $\delta_g \lambda_H < 0$). This in turn implies that the gauge
contribution~(\ref{eq:CW_g})--(\ref{eq:STrM4_gH}) is not enough
to bring $v$ to its true value. Conversely, one may adjust the
parameters of the model so that the correct value of the electroweak
scale is obtained, but then the radiative corrections to $\lambda_H$
are too small and the Higgs mass falls below the LEP bound.
This means that the model must be extended to be fully realistic.
A first way to do so is to add an $SU(3)_2$-breaking sector that
generates a sizeable tree-level quartic coupling $\lambda_0$,
as was done in a different $SU(3)$ model in Ref.~\cite{Roy05}.
Then large corrections from the (s)top sector are no longer needed
to satisfy the LEP bound, and the electroweak scale is obtained
with little fine-tuning. Another possibility is to invoke some additional
convex contribution $\delta_{\rm extra} m^2_H\, |H|^2$ to the effective
Higgs potential (presumably arising from loops involving the heavy gauge
and Higgs fields, or from some extra $SU(3)_2$-breaking sector
to be added to the model) in order to obtain the proper value of the
Higgs VEV. In this case large corrections from the (s)top sector
to $\lambda_H$ are still needed to satisfy the LEP constraint
and the fine-tuning is more significant.

\begin{figure}[h!]
\begin{minipage}[b]{0.47\linewidth}
\centering
\includegraphics[width=6 cm]{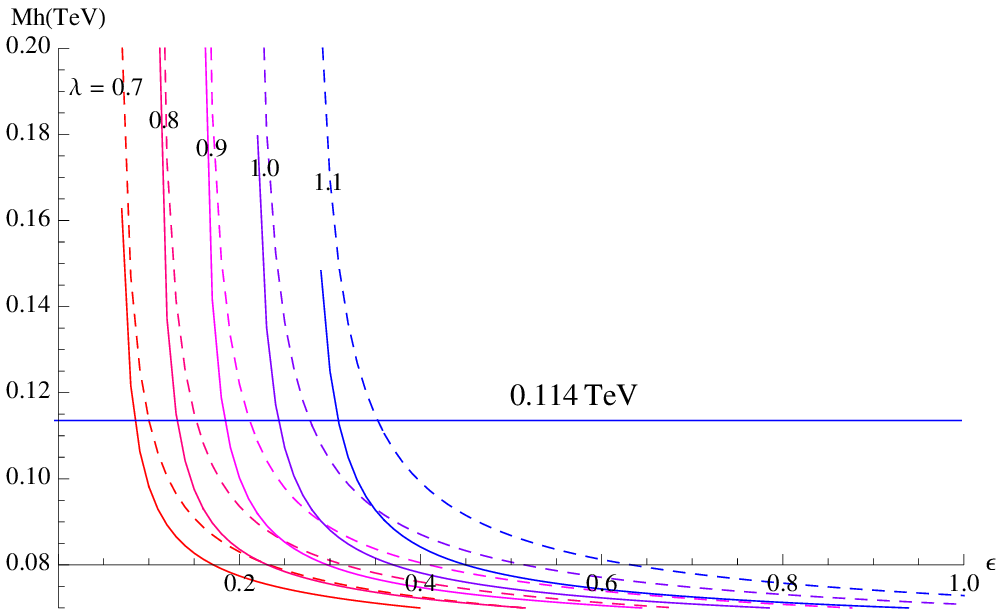}
\caption{Higgs boson mass as a function \newline
of $\lambda$ and $\epsilon$, for $M=500\TeV$,
$m_{\rm soft} \sim 1\TeV$, \newline
$y_{1}=1$, $F=7\TeV$ and $\xi=0.00014$ [in TeV].}
\label{fig:Mh}
\end{minipage}
\hspace{0.5cm}
\begin{minipage}[b]{0.47\linewidth}
\centering
\includegraphics[width=6 cm]{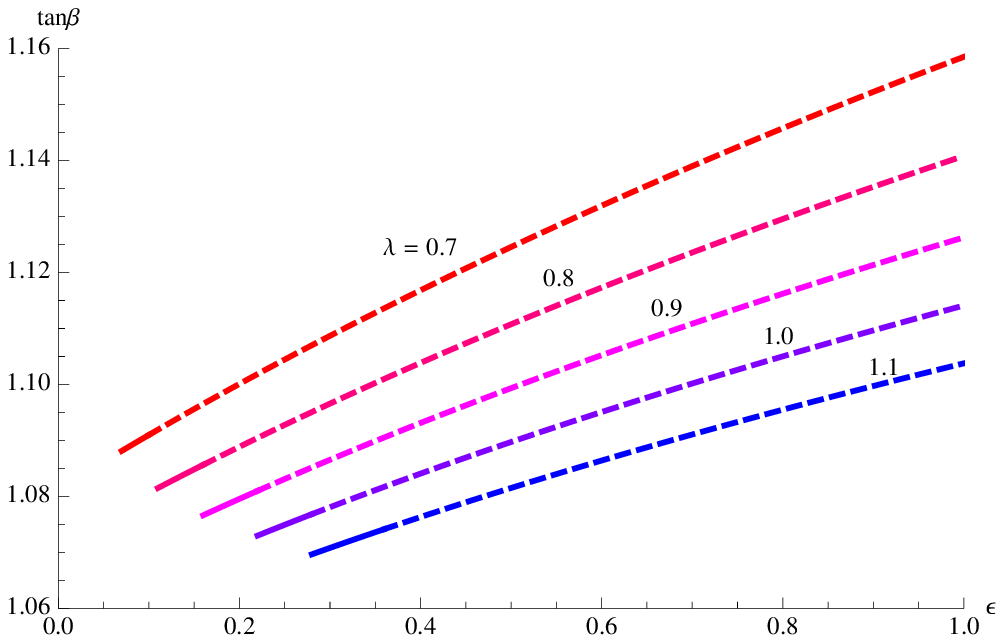}
\caption{$\tan\beta$ as a function of $\lambda$ and $\epsilon$. \newline
Other parameters chosen as in Fig.~\ref{fig:Mh}.}
\label{fig:tanbeta}
\end{minipage}
\end{figure}

\begin{figure}[h!]
\begin{minipage}[b]{0.47\linewidth}
\centering
\includegraphics[width=6 cm]{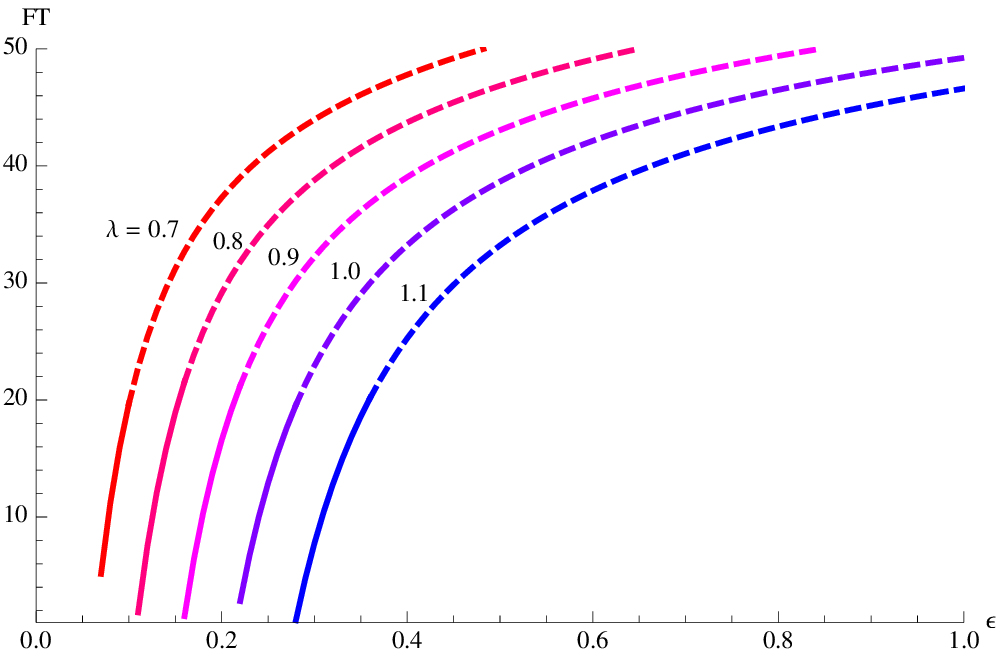}
\caption{Fine-tuning as a function of $\lambda$ and $\epsilon$.
Other parameters chosen as in Fig.~\ref{fig:Mh}.}
\label{fig:FT}
\end{minipage}
\hspace{0.5cm}
\begin{minipage}[b]{0.47\linewidth}
\centering
\includegraphics[width=6 cm]{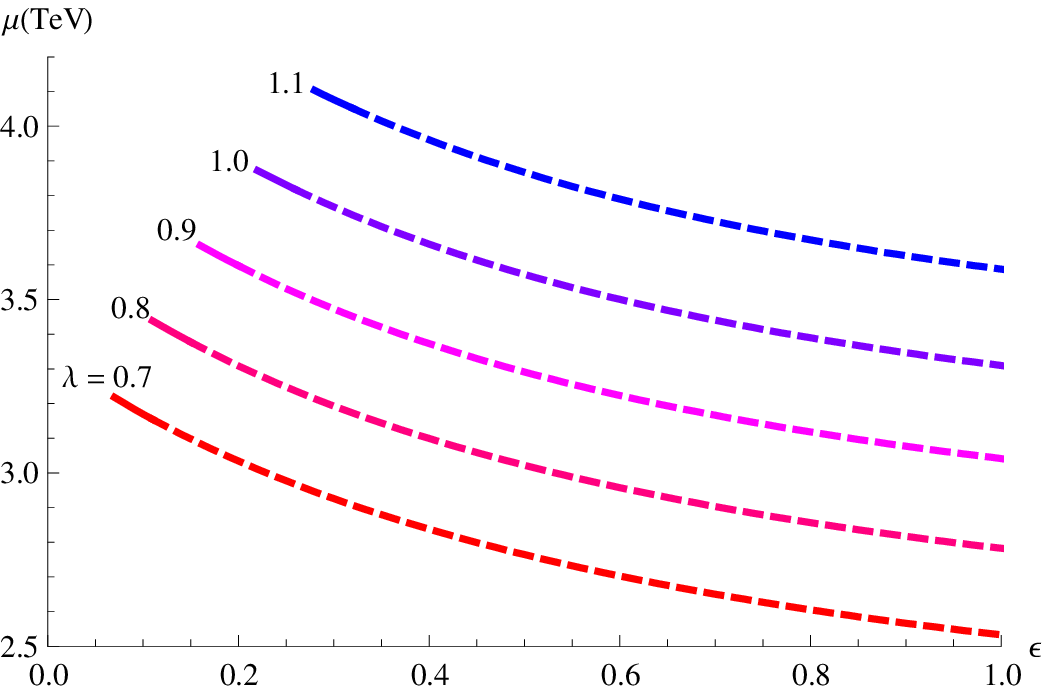}
\caption{$\mu=-\lambda v_{N}$ as a function of $\lambda$ and $\epsilon$. \newline
Other parameters chosen as in Fig.~\ref{fig:Mh}.}
\label{fig:mu}
\end{minipage}
\end{figure}

Let us investigate the second possibility.
To fix the size of the (s)top and gauge contributions,
we require that the resulting $\delta \lambda_H$ be large enough
to satisfy the LEP bound for a mass of the
$SU(3)$ top partner $m_T$ in the 1--10 TeV range. In practice,
the parameter values chosen in Fig.~\ref{fig:f_xi-00014}
for the spontaneous breaking of the global $SU(3)_2$ symmetry
turn out to be convenient for that purpose and we adopt them in our numerical study.
Then we adjust the extra contribution $\delta_{\rm extra} m^2_H\, |H|^2$
to obtain the correct value of the electroweak scale.
Let us now present the numerical results for the same choice
of parameters as in Fig.~\ref{fig:f_xi-00014} (all coupling values in the figures
are given at the messenger scale). The value
of the Higgs mass $M_h \simeq 2 \sqrt{\lambda_0 + \delta \lambda_H}\, v$
is displayed in Fig.~\ref{fig:Mh} (the dashed lines show the Higgs mass
predicted by the (s)top sector contribution to the effective potential alone),
while the corresponding value of $\tan\beta$ is shown in Fig.~\ref{fig:tanbeta}.
The curves in Figs.~\ref{fig:tanbeta} to~\ref{fig:mu} are dashed in the
region of the parameter space where the Higgs mass lies below
the LEP bound.

The source of fine-tuning in the model lies in the large
radiative corrections to $\lambda_H$ from the (s)top sector that are
required in order to satisfy the LEP bound. This in turn implies a large
and negative $\delta_t m^2_H$ that must be compensated for
by $\delta_g m^2_H$ and $\delta_{\rm extra} m^2_H$ so as to
obtain the proper value of the electroweak scale $v = 174\GeV$.
One can estimate this fine-tuning with the following quantity:
\begin{equation}
  FT\, =\, \left|\frac{|\delta_t m_{H}^{2}|-\left|\delta m_{H}^{2}\right|}
  {\delta m_{H}^{2}}\right|\, ,
\label{FT}
\end{equation}
where $\delta m_{H}^{2} = \delta_t m_{H}^{2} + \delta_g m_{H}^{2}
+ \delta_{\rm extra} m_{H}^{2}$ is the mass squared parameter in the
Higgs potential~(\ref{ex}).
The numerical results for the fine-tuning parameter $FT$ are
presented in Fig.~\ref{fig:FT}. Figs.~\ref{fig:Mh} to~\ref{fig:FT} show that
successful electroweak symmetry breaking with a Higgs boson
mass above the LEP bound and a fine-tuning around $FT\sim20$
can be achieved for $\lambda\sim1$, $\epsilon\sim0.3$ and
$\xi\sim10^{-4}$ 
(corresponding to $f \sim1\TeV$).

\subsection{Physical spectrum}

Let us finally discuss the physical spectrum of the model.
The gross features of the Higgs spectrum are the following
(omitting the heavy fields $\Phi_U$, $\Phi_D$ and $N'$,
which have masses of order $F$, out of reach of the LHC).
The spontaneous breaking of the $SU(3)_2$ symmetry yields
a massive Higgs doublet
$H' \sim \cos \beta \left( i \sigma^2 H^*_u\right) - \sin \beta H_d$,
which describes a CP-even and a CP-odd neutral scalars
as well as a charged one, all with the same tree-level mass
$\sqrt{2\mu^2 + m^2_u + m^2_d} \approx \sqrt{2}\, |\mu|$,
hence in the multi-TeV range (see Fig.~\ref{fig:mu}).
In the singlet sector, we have 4 heavy scalars with masses
of order a few $\mu$ and a lighter one with a mass of order
$\lambda f \lesssim 1 \TeV$; the remaining singlet $\eta$ is a
pseudo-Goldstone boson and gets a small mass at the one-loop
level~\cite{Bellazzini0906}.
Apart from this singlet, whose phenomenology has been studied
in Refs.~\cite{Berezhiani05,Bellazzini0906,Bellazzini0910},
the Higgs sector contains a single light state with
Standard Model-like properties.

The higgsinos (both doublet and singlets) also have large masses
of order $\mu$. The rest of the superpartner spectrum is
representative of gauge-mediated models with a low messenger
scale. The $SU(3)$ partner of the top quark has a mass
$m_T \simeq y_1 F \approx 7\TeV$ in the region of parameter
space considered and is not accessible at the LHC, similarly
to the heavy gauge bosons associated with the broken
$SU(3)_W \times U(1)_X$ generators.

\section{Conclusions}
\label{sec:conclusions}

In this paper, we studied the interplay between the spontaneous
breaking of a global symmetry of the Higgs sector and gauge-mediated
supersymmetry breaking, in the framework of a supersymmetric model
with global $SU(3)$ symmetry.
In addition to solving the supersymmetric flavour problem and
alleviating the little hierarchy problem by identifying the Higgs
boson with a pseudo-Goldstone boson, this scenario
presents several advantages.

First, gauge mediation provides
a mechanism for breaking  the global symmetry protecting
the Higgs mass, namely through the loop-induced tadpole of an
$SU(3)$ singlet scalar field. 
A non-trivial success of the model studied in this paper,
compared with previous attempts in the literature,
is to ensure that the global symmetry
breaking vacuum is indeed the global minimum of the scalar
potential, and the possibility to control the shape of the potential
by varying the tadpole scale is instrumental in this.

Second, the global symmetry provides an elegant solution to the
$\mu$/$B\mu$ problem of gauge mediation. Much like in the NMSSM,
the $\mu$ and $B\mu$ parameters are generated by the VEVs
of the scalar and F-term components of a singlet superfield,
but the global $SU(3)$ symmetry ensures
that the relation $(B\mu)^2 = (|\mu|^2 + m^2_u)(|\mu|^2 + m^2_d)$
is satisfied at the minimum of the tree-level scalar potential, implying
that the electroweak scale is insensitive to the actual value
of the $\mu$ parameter.
As a result the $\mu$ parameter, which sets the scale of the heavy
Higgs masses, may be large without creating a strong
fine-tuning in the Higgs potential.

Finally, the combined effect of supersymmetry and of the global
symmetry ensures a ``double protection'' of the
Higgs potential, allowing for a reduced fine-tuning with respect
to the MSSM. We computed the one-loop corrections to the potential
of the pseudo-Goldstone Higgs boson coming from the (s)top
and gauge-Higgs sectors, and checked that they indeed trigger
electroweak symmetry breaking.
However, the specific model studied in this paper has
a suppressed tree-level quartic Higgs coupling and fails to bring
the Higgs mass above the LEP bound.
We showed that an additional contribution
$\delta_{\rm extra} m^2_H\, |H|^2$ to the Higgs potential,
arising from some extra $SU(3)$-breaking sector,
can solve this problem  with
a moderate fine-tuning of order $1/20$. Alternatively,
one may try to generate a sizeable tree-level quartic Higgs coupling
along the lines of Ref.~\cite{Roy05}.

The model predicts a rather low messenger scale, a small $\tan\beta$
value, a light Higgs boson with Standard Model-like properties,
and heavy higgsinos.

\vskip .3cm


\section*{Acknowledgments}
The work of AK was supported by the MNiSzW scientific research
grant N N202 103838 (2010 - 2012).
The work of SL was partially supported by the European Community
under the contracts MTKD-CT-2005-029466 and PITN-GA-2009-237920.

\begin{appendix}

\renewcommand{\theequation}{A.\arabic{equation}}
\setcounter{equation}{0}  

\section{Renormalization Group Equations}
\label{app:RGEs}

In this appendix, we give the renormalization group equations
(RGEs) valid between the messenger scale $M$ and the $SU(3)_W$
breaking scale $F$ for all relevant superpotential parameters and
soft terms. For convenience, we recall their definition below:
\begin{align}
W\, \ni\, &\ \lambda'N'\left(\Phi_{U}\Phi_{D}-\mu^{2}\right)
  +\lambda{}N{\cal H}_{u}{\cal H}_{d}+\frac{\kappa}{3}N^{3}
  +y_{1}\Phi_{U}\Psi_{Q}T^{c}+y_{2}{\cal H}_{u}\Psi_{Q}t^{c} , \\
V_{\rm soft}\, \ni\, &\ M^{2}_{U}\left|\Phi_{U}\right|^{2}
+M^{2}_{D}\left|\Phi_{D}\right|^{2}+M^{2}_{N'}\left|N'\right|^{2}
+m_{u}^{2}\left|{\cal H}_{u}\right|^{2}+m_{d}^{2}\left|{\cal H}_{d}\right|^{2}
+m_{N}^{2}\left|N\right|^{2}
\nonumber \\
&\, +m_{\Psi_{Q}}^{2}\left|\Psi_{Q}\right|^{2}
+m_{t^{c}}^{2}\left|m_{t^{c}}\right|^{2}+m_{T^{c}}^{2}\left|m_{T^{c}}\right|^{2}
+\left(\lambda'A_{\lambda'}N'\Phi_{U}\Phi_{d} \right.
\nonumber \\
&\, \left. +\lambda{}A_{\lambda}N{\cal H}_{u}{\cal H}_{d}
+\frac{\kappa}{3}A_{\kappa}N^{3}+y_{1}A_{y_{1}}\Phi_{U}\Psi_{Q}T^{c}
+y_{2}A_{y_{2}}{\cal H}_{u}\Psi_{Q}t^{c}+\mbox{h.c.}\right) .
\label{eq:soft_terms}
\end{align}
In the RGEs below, $g_C$, $g_W$ and $g_X$ are the $SU(3)_C$,
$SU(3)_W$ and $U(1)_X$ gauge couplings, respectively;
$M_C$, $M_W$ and $M_X$ are the associated gaugino masses;
and $t \equiv (1/16\pi^{2}) \ln\mu$.

\begin{align}
\frac{d}{dt}\, \lambda'\, =\, &\ \lambda'\left(5\lambda'^{2}+3y_{1}^{2}
-\frac{16}{3}g_{W}^{2}-\frac{4}{9}g_{X}^{2}\right) , \\
\frac{d}{dt}\, \lambda\, =\, &\ \lambda\left(5\lambda^{2}+2\kappa^{2}
+3y_{2}^{2}-\frac{16}{3}g_{W}^{2}-\frac{4}{9}g_{X}^{2}\right) , \\
\frac{d}{dt}\, \kappa\, =\, &\ 3\kappa\left(3\lambda^{2}+2\kappa^{2}\right) ,
\label{k} \\
\frac{d}{dt}\, y_{1}\, =\, &\ y_{1}\left(7y_{1}^{2}+y_{2}^{2}
+\lambda'^{2}-\frac{16}{3}g_{C}^{2}-\frac{16}{3}g_{W}^{2}
-\frac{4}{3}g_{X}^{2}\right) , \\
\frac{d}{dt}\, y_{2}\, =\, &\ y_{2}\left(y_{1}^{2}+7y_{2}^{2}
+\lambda^{2}-\frac{16}{3}g_{C}^{2}-\frac{16}{3}g_{W}^{2}
-\frac{4}{3}g_{X}^{2}\right) ,
\label{y} \\
\frac{d}{dt}\, A_{\lambda'}\, =\, &\ 10\lambda'^{2}A_{\lambda'}
+6y_{1}^{2}A_{y_{1}}-\frac{32}{3}g_{W}^{2}M_{W}
-\frac{8}{9}g_{X}^{2}M_{X}\, , \\
\frac{d}{dt}\, A_{\lambda}\, =\, &\ 10\lambda^{2}A_{\lambda}
+4\kappa^{2}A_{\kappa}+6y_{2}^{2}A_{y_{2}}
-\frac{32}{3}g_{W}^{2}M_{W}-\frac{8}{9}g_{X}^{2}M_{X}\, ,
\label{Al} \\
\frac{d}{dt}\, A_{\kappa}\, =\, &\ 6\left(3\lambda^{2}A_{\lambda}
+2\kappa^{2}A_{\kappa}\right) ,
\label{Ak} \\
\frac{d}{dt}\, A_{y_{1}}\, =\, &\ 14y_{1}^{2}A_{y_{1}}
+2\lambda'^{2}A_{\lambda'}-\frac{32}{3}g_{C}^{2}M_{C}
-\frac{32}{3}g_{W}^{2}M_{W}-\frac{8}{3}g_{X}^{2}M_{X}\, , \\
\frac{d}{dt}\, A_{y_{2}}\, =\, &\ 14y_{2}^{2}A_{y_{2}}
+2\lambda^{2}A_{\lambda}-\frac{32}{3}g_{C}^{2}M_{C}
-\frac{32}{3}g_{W}^{2}M_{W}-\frac{8}{3}g_{X}^{2}M_{X}\, ,
\label{Ay} \\
\frac{d}{dt}\, m^{2}_{U}\, =\, &\ 6y^{2}_{1}\left(m^{2}_{U}
+m^{2}_{\Psi_{Q}}+m^{2}_{T^{c}}+A^{2}_{y_{1}}\right)
+2\lambda'^{2}\left(m_{U}^{2}+m^{2}_{D}+m^{2}_{N'}+A^{2}_{\lambda'}\right)
\nonumber \\
&\, -\frac{32}{3}g_{W}^{2}M_{W}^{2}-\frac{8}{9}g_{X}^{2}M_{X}^{2}\, , \\
\frac{d}{dt}\, m^{2}_{D}\, =\, &\ 2\lambda'^{2}\left(m_{U}^{2}
+m^{2}_{D}+m^{2}_{N'}+A^{2}_{\lambda'}\right)
-\frac{32}{3}g_{W}^{2}M_{W}^{2}-\frac{8}{9}g_{X}^{2}M_{X}^{2}\, , \\
\frac{d}{dt}\, m^{2}_{N'}\, =\, &\ 6\lambda'^{2}\left(m_{U}^{2}
+m^{2}_{D}+m^{2}_{N'}+A^{2}_{\lambda'}\right) , \\
\frac{d}{dt}\, m^{2}_{u}\, =\, &\ 6y^{2}_{2}\left(m^{2}_{u}
+m^{2}_{\Psi_{Q}}+m^{2}_{t^{c}}+A^{2}_{y_{2}}\right)
+2\lambda^{2}\left(m_{u}^{2}+m^{2}_{d}+m^{2}_{N}+A^{2}_{\lambda}\right)
\nonumber \\
&\, -\frac{32}{3}g_{W}^{2}M_{W}^{2}-\frac{8}{9}g_{X}^{2}M_{X}^{2}\, ,
\end{align}
\begin{align}
\frac{d}{dt}\, m^{2}_{d}\, =\, &\ 2\lambda^{2}\left(m_{u}^{2}
+m^{2}_{d}+m^{2}_{N}+A^{2}_{\lambda}\right)
-\frac{32}{3}g_{W}^{2}M_{W}^{2}-\frac{8}{9}g_{X}^{2}M_{X}^{2}\, ,
\label{md} \\
\frac{d}{dt}\, m^{2}_{N}\, =\, &\ 6\lambda^{2}\left(m_{u}^{2}
+m^{2}_{d}+m^{2}_{N}+A^{2}_{\lambda}\right)
+4\kappa^{2}\left(3m_{N}^{2}+A_{\kappa}^{2}\right) ,
\label{mn} \\
\frac{d}{dt}\, m^{2}_{\Psi_{Q}}\, =\, &\ 2y_{1}^{2}\left(m^{2}_{U}
+m_{\Psi_{Q}}^{2}+m^{2}_{T^{c}}+A^{2}_{y_{1}}\right)
+2y_{2}^{2}\left(m^{2}_{u}+m_{\Psi_{Q}}^{2}+m^{2}_{t^{c}}+A^{2}_{y_{2}}\right)
\nonumber \\
&\, -\frac{32}{3}g_{C}^{2}M_{C}^{2}-\frac{32}{3}g_{W}^{2}M_{W}^{2}
-\frac{8}{9}g_{X}^{2}M_{X}^{2}\, ,
\label{mp} \\
\frac{d}{dt}\, m^{2}_{t^{c}}\, =\, &\ 6y_{2}^{2}\left(m^{2}_{u}
+m_{\Psi_{Q}}^{2}+m^{2}_{t^{c}}+A^{2}_{y_{2}}\right)
-\frac{32}{3}g_{C}^{2}M_{C}^{2}-\frac{32}{9}g_{X}^{2}M_{X}^{2}\, ,
\label{mt} \\
\frac{d}{dt}\, m^{2}_{T^{c}}\, =\, &\ 6y_{1}^{2}\left(m^{2}_{U}
+m_{\Psi_{Q}}^{2}+m^{2}_{T^{c}}+A^{2}_{y_{1}}\right)
-\frac{32}{3}g_{C}^{2}M_{C}^{2}-\frac{32}{9}g_{X}^{2}M_{X}^{2}\, .
\end{align}
%


\renewcommand{\theequation}{B.\arabic{equation}}
\setcounter{equation}{0}  

\section{Values of soft terms at the messenger scale}
\label{app:soft_terms}

In this appendix, we give the boundary conditions for the soft terms
in Eq.~(\ref{eq:soft_terms}) at the messenger scale $M$.
\begin{align}
m_{U}^{2}\, =\, &\ m_{D}^{2}\, =\, \frac{1}{(16\pi^{2})^{2}}
\left( \frac{8}{3}g_{W}^{4}+\frac{8}{27}g_{X}^{4} \right) \Lambda^{2}\, , \\
m_{u}^{2}\, =\, &\ m_{d}^{2}\, =\, \frac{1}{(16\pi^{2})^{2}}
\left( \frac{8}{3}g_{W}^{4}+\frac{8}{27}g_{X}^{4}-6\lambda^{2}\xi^{2} \right)
\Lambda^{2}\, , \\
m_{N'}^{2}\, =\, &\ 0\ , \\
m_{N}^{2}\, =\, &\ \frac{1}{(16\pi^{2})^{2}} \left( 48\xi^{4}
-24\kappa^{2}\xi^{2}-16g_{C}^{2}\xi^{2}-16g_{W}^{2}\xi^{2}
-\frac{8}{3}g_{X}^{2}\xi^{2} \right) \Lambda^{2}\, , \\
m_{\Psi_{Q}}^{2}\, =\, &\ \frac{1}{(16\pi^{2})^{2}} \left( \frac{8}{3}g_{C}^{4}
+\frac{8}{3}g_{W}^{4}+\frac{8}{27}g_{X}^{4} \right) \Lambda^{2}\, , \\
m_{t^{c}}^{2}\, =\, &\ \frac{1}{(16\pi^{2})^{2}} \left( \frac{8}{3}g_{C}^{4}
+\frac{34}{27}g_{X}^{4} \right) \Lambda^{2}\, , \\
m_{T^{c}}^{2}\, =\, &\ \frac{1}{(16\pi^{2})^{2}} \left( \frac{8}{3}g_{C}^{4}
+\frac{34}{27}g_{X}^{4} \right) \Lambda^{2}\, , \\
A_{\lambda}\, =\, &\ \frac{A_{\kappa}}{3}\,
=\, -\frac{6\xi^{2}}{16\pi^2}\, \Lambda\, , \\
A_{\lambda'}\, =\, &\ A_{y_{1}}\, =\, A_{y_{2}}\, =\, 0\ .
\end{align}
%


\renewcommand{\theequation}{C.\arabic{equation}}
\setcounter{equation}{0}  

\section{Renormalization Group Equations for ``split''
superpotential couplings}
\label{app:RGEs_split}

Below the $SU(3)_W$ breaking scale $F$, the $SU(3)_2$-invariant
superpotential couplings $\lambda$ and $y_2$ are split into separate
couplings for the doublet and singlet components of the Higgs
triplets ${\cal H}_u$ and ${\cal H}_d$:
\begin{equation}
W\, \ni\, \lambda_{s}NS_{u}S_{d}+\lambda_{d}NH_{u}H_{d}
+y_{2,s}S_{u}Tt^{c}+y_{2,d}H_{u}Qt^{c}\, .
\end{equation}
The associated RGEs, valid below the scale $F$, are given by:
\begin{align}
\frac{d}{dt}\, \lambda_{s}\, =\, &\ \lambda_{s}\left(3y_{2,s}^{2}+3\lambda_{s}^{2}+2\lambda_{d}^{2}+2\kappa^{2}\right) , \\
\frac{d}{dt}\, \lambda_{d}\, =\, &\ \lambda_{d}\left(3y_{2,d}^{2}+\lambda_{s}^{2}+4\lambda_{d}^{2}+2\kappa^{2}-2\left(\frac{1}{2}g'^{2}+\frac{3}{2}g^{2}\right)\right) , \\
\frac{d}{dt}\, y_{2,s}\, =\, &\ y_{2,s}\left(5y_{2,s}^{2}+2y_{2,d}^{2}+\lambda_{s}^{2}-2\left(\frac{8}{9}g'^{2}+\frac{8}{3}g_{C}^{2}\right)\right) , \\
\frac{d}{dt}\, y_{2,d}\, =\, &\ y_{2,d}\left(y_{2,s}^{2}+6y_{2,d}^{2}+\lambda_{d}^{2}-2\left(\frac{13}{18}g'^{2}+\frac{3}{2}g^{2}+\frac{8}{3}g_{C}^{2}\right)\right) ,
\end{align}
where $g_C$, $g$ and $g'$ are the $SU(3)_C$, $SU(2)_L$ and $U(1)_Y$
gauge couplings, respectively, and $t \equiv (1/16\pi^{2}) \ln\mu$.
The matching conditions between the $SU(3)_W \times U(1)_X$ and
$SU(2)_L \times U(1)_Y$ gauge couplings at the scale $F$ read:
\begin{equation}
g\, =\, g_W\, , \qquad \qquad
g'\, =\, \frac{g_W g_X }{\sqrt{g^2_W + g^2_X/3}}\ .
\end{equation}
For completeness, we also give the matching conditions for the
gaugino masses:
\begin{equation}
M_2\, =\, M_W\, , \qquad \qquad
M_1\, =\, \frac{g^2_X M_W + 3g^2_W M_X}{3g^2_W + g^2_X}\ .
\end{equation}

\end{appendix}



\end{document}